\documentstyle[11pt]{article}
\newcommand{\be}{\begin{equation}}
\newcommand{\ee}{\end{equation}}
\newcommand{\ba}{\begin{eqnarray}}
\newcommand{\ea}{\end{eqnarray}}

\newcommand{\bea}{\begin{array}{l}}
\newcommand{\eea}{\end{array}}
\newcommand{\ds}{\displaystyle}

\newcommand{\infu}{\int {\cal D}[V] \int \prod_{i=1}^{3} {\cal D}[\varphi_i] 
\, {\cal D} [\bar{\varphi}_i]}
\newcommand{\mdm}{M \partial_{M}}
\newcommand{\ie}{{\it i.e.}}
\newcommand{\inffu}{\int \prod_{i=1}^{3} {\cal D}[\varphi_i] 
\, {\cal D} [\bar{\varphi}_i]}
\newcommand{\ifu}{\int \prod_{i=1}^{3} {\cal D}[\varphi'_i] 
\, {\cal D} [\bar{\varphi}'_i]}
\newcommand{\innffu}{\int \prod_{i=1}^{3} {\cal D}[\varphi^{''}_i] 
\, {\cal D} [\bar{\varphi}^{''}_i]}

\begin{document}

\title{{\bf \huge{Exact renormalization group equation in presence of rescaling anomaly.}}}
\author{S. Arnone\thanks{e-mail: Arnone@roma1.infn.it},
C. Fusi\thanks{e-mail: Fusic@roma1.infn.it} 
and K. Yoshida\thanks{e-mail: Yoshida@roma1.infn.it}\\
Dipartimento di Fisica, Universit\`a di Roma ``La Sapienza''\\
Piazzale Aldo Moro, 2 - 00185 Roma\\
and\\
I.N.F.N, Sezione di Roma I\\}
\maketitle
\begin{abstract}
Wilson'~s approach to renormalization group is reanalyzed for supersymmetric 
Yang-Mills theory.
Usual demonstration of exact renormalization group equation 
must be modified due to the presence of the so called Konishi anomaly 
under the rescaling of superfields.\\
We carry out the explicit computation for $N=1$ SUSY Yang-Mills theory with the
simpler, gauge invariant regularization method, recently proposed by 
Arkani-Hamed and Murayama.\\
The result is that the Wilsonian action $S_M$ consists of two terms, {\it i.e.}
 the non anomalous term, which obeys Polchinski'~s flow equation 
and Fujikawa-Konishi determinant contribution. This latter 
is responsible for Shifman-Vainshtein relation of exact $\beta$-function.\\
\end{abstract}
\newpage
\section{Introduction.}
In two interesting papers~\cite{mu1,mu2} Arkani-Hamed and Murayama have 
analyzed the problem of renormalization group (RG) invariance of the 
so called exact results in SUSY gauge field theory.  
They have shown convincingly the mechanism by which Shifman - Vainshtein'~s
results on $\beta$-function for $N=1$ SUSY Yang-Mills (SYM) theory come out. 
In~\cite{mu2} they have proposed an explicit regularization method for such a 
theory which is free of the usual problem of the inconsistency with local gauge
 symmetry~\cite{Bec,Mor}. Although this method has its own problems seeing 
that it depends on $N=4$ SYM, it permits the great simplification in 
theoretical 
investigation such as the derivation of exact renormalization group equation 
(ERGE) (assuming that the regularization through ``finite'' $N=4$ theory is 
basically sound).\\
In this note, we reformulate the results of~\cite{mu1,mu2} in the spirit 
of ERGE 
due to Polchinski~\cite{Po}, which is the formal way to apply Wilson'~s 
decimation method to the continuum field theory~\cite{Wi}.\\
We note, first of all, that the vacuum energy of auxiliary chiral superfields 
(belonging to $N=4$ theory) in the presence of external vector fields, being 
in principle dependent on the relevant gauge fields, cannot be discarded 
as in~\cite{Po}. We apply the known result on rescaling 
anomaly~\cite{Kon} for chiral superfields to evaluate explicitly such a vacuum
contribution.\\
This leads to the conclusion that the Wilsonian action $S_M$, $M$ being a cutoff mass, consists of two parts, the normal term which obeys the RG flow equation
of the type proposed in~\cite{Po} and Fujikawa-Konishi determinant 
contribution~\cite{Fuji,Kon}; it is the last term which is responsible for the 
Shifman-Vainshtein result on the $\beta$-function.
These results are exact and they neither depend on the perturbation expansion (
except for the calculation of anomaly which is, however, immune to the 
radiative corrections), nor on the large $M$ approximation.\\
\section{Regularization of $N=1$ SYM theory and ERG equation.}
\label{sec:2}
\subsection{Regularized $N=1$ SYM.}
We have to regularize the $N=1$ SYM theory whose classical action is
\be \label{hol}
S(V) = \frac{1}{16} \int d^4 x \, d^2 \theta \frac{1}{g^2_{h}} W^{\alpha}_a 
W_{\alpha}^a + h.c.
\ee
This is the so called holomorphic representation, which is written in a 
manifestly holomorphic way in the gauge coupling 
\be
\frac{1}{g^2_{h}} = \frac{1}{g^2} + \frac{i \theta}{8 \pi^2}
\ee
(throughout the present note, we 
adopt the conventions used in~\cite{mu1,mu2,WB}).\\		
The more conventional ``canonical representation''
\be
S_c(V_c)  = \frac{1}{16} \int d^4 x \, d^2 \theta \left( \frac{1}{g^2_{c}} 
+ \frac{i \theta}{8 \pi^2} \right) W^{\alpha}_a (g_c V_c) W_{\alpha}^a (g_c 
V_c) + h.c.
\ee
can be obtained from Eq. (\ref{hol}) by the rescaling transformation. 
For details, see~\cite{mu2}.\\
To quantize Eq. (\ref{hol}) with suitable cut-off, we follow~\cite{mu2} and start
from the $N=4$ SYM in $4 D$ which is believed to be finite~\cite{24}. Apart from the 
ghosts necessary for gauge invariance, this latter consists of 
vector superfield 
$V$ and chiral and anti-chiral triplets $(\varphi_i, \bar{\varphi}_i)_{i=
1,2,3}$ ($\bar{D}_{\dot{\alpha}} \varphi_i = 0$).\\
All the superfields are in the adjoint representation of the gauge group $G$.\\
The classical action is
\be \label{N4}
\bea
{\ds S_{N=4} (V,\varphi,\bar{\varphi}; g) = \frac{1}{16} \int d^4 x \, d^2 
\theta \frac{1}{g^2} W^a_{\alpha} W^{\alpha}_a +
h.c. +\int dx \, d^4 \theta  \, {\cal R}e  \left( \frac{2}{g^2}\right) t_2(A) 
\bar{\varphi}^i e^V \varphi^i +}\nonumber\\[0.35cm]
{\ds \int d^4 x \, d^2 \theta \, {\cal R}e  \left( \frac{1}{g^2} \right)
\sqrt{2} \, Tr \left( \varphi^i \, [\varphi^j,\varphi^k] \right) 
\frac{\epsilon^{ijk}}{3!} + h.c. }\\
\eea
\ee
where $t_2 (A)$ is the Dynkin index of the adjoint representation of $G$.\\
Following~\cite{mu2} we write down the regulated $N=1$ SYM theory as
\be \label{Z0}
\bea
{\ds Z_0(J_i,\bar{J}_i,J_V) = \infu \exp i \left[ S_{N=4} (V,\varphi,
\bar{\varphi}; g_0)+ \frac{M_0}{2} \int  
\varphi_i^2 +\right. }\nonumber\\
{\ds \left. + h.c. + \int J_i \varphi_i + h.c. + \int J_V V \right]}\\
\eea
\ee
where ${\ds \int \varphi_i^2}$ stands for ${\ds \int d^4 p \, d^2 \theta \, 
\sum_{a,i} \varphi_i^a(-p) \varphi_i^a (p)}$.\\
This expression must be really supplemented with appropriate gauge fixing
term and corresponding ``ghosts''. But as far as our regularization is 
concerned, these terms and auxiliary fields do not influence the arguments in
any essential way. This is not the case in more conventional cut-off 
method~\cite{Po,Bec,Mor}.\\
The mass term
\be
\frac{M_0}{2} \int d^4 x \, d^2 \theta \, \varphi_i^2 + h.c.
\ee
breaks $N=4$ SUSY but preserves the finiteness of the RHS of Eq. (\ref{Z0}) 
(soft breaking~\cite{Parkes}). This term is gauge invariant\footnote{ 
This expression differs from the one in~\cite{mu2} by the 
Dynkin index of the adjoint representation of the gauge group.}.\\
From a physical point of view, one takes large enough $M_0$ with respect to the momentum scale $p \sim M'$ which characterizes low energy physics. Thus, at
$p \sim M' \ll M_0$, it is expected that the influences of auxiliary 
superfields $\varphi_i$ and $\bar{\varphi}_i$ are negligible.
\subsection{ERG equation.}
The first step for applying Wilson'~s idea of RG is to vary $M_0$ to $M<M_0$ 
and search for the new action functional (Wilsonian action) $S_M (V,\varphi_i,
\bar{\varphi}_i)$ in such a way that the physics is unchanged lowering the
regularizing mass, \ie
\be \label{ZM}
\bea
{\ds Z_M(J_i,\bar{J}_i,J_V) = \infu \exp i \left[ S_M (V,\varphi,
\bar{\varphi}) + \frac{M}{2} \int  
\varphi_i^2 +\right. }\nonumber\\
{\ds \left. + h.c. + \int f(M) J_i \, \varphi_i + h.c. + \int J_V V \right] }\\
\eea
\ee
must be equal to the original $Z_0$ (Eq. (\ref{Z0})).\\
$f(M)$ is the renormalization factor to be determined.\\
We derive the equation which should be obeyed by $S_M$ following Polchinski~\cite{Po}. Let us write down the ``trivial'' equality
\be \label{treq}
\bea
{\ds 0 = \int d^4 p \, \mdm C(p,M) \infu \left\{ \int d^2 \theta
\frac{\delta}{\delta \varphi_i(p)} \bigg[ -i M \varphi_i(p)+
\right. }\nonumber\\[0.35cm]
{\ds \left. \frac{1}{2} \frac{\delta}{\delta \varphi_i(-p)} \bigg] + h.c. 
\right\} \exp \left(i S_{tot} \right) }\\
\eea
\ee
with ${\ds S_{tot} \doteq S_M + \frac{M}{2} \int dp \, 
\varphi_i^2 + h.c. \int f(M) J_i \varphi_i + h.c. +\int J_V V}$.\\
The cut-off function $C(p,M)$ is simply equal to $M^{-1}$ in our case.\\
To go further, however, we must decide what to do with the singular term
\be \label{site}
\delta_{ii} \doteq \sum_{i,a} \frac{\delta \varphi_i^a (p) }{\delta 
\varphi_i^a (p).}
\ee
In~\cite{Po}, the similar term is discarded because it adds only a constant
(perhaps divergent) over all factor to $Z_M$.\\
In our case, apart from being singular, it may depend on $V$ superfield which
does not enter directly in the regularization procedure.\\
In other words, one cannot ignore the vacuum energy of $ \varphi_i $ and
$ \bar{\varphi}_i$ superfields in external $ V $ superfield since it 
depends, in general, on $ V $.\\
Thus it is necessary to analyze the quantity $\delta_{ii}$ and find its exact
value within our regularization scheme.\\
To this end, we apply the result on rescaling anomaly studied by Konishi and 
Schizuya~\cite{Kon}. Under the (infinitesimal) rescaling transformation of 
chiral superfields $\varphi_i$ in the presence of external gauge superfield $V$
\be \label{resc}
\varphi_i = e^{\delta \alpha} \varphi'_i
\ee
(here one must consider $\delta \alpha (x)$ as a chiral superfield)\\
one has a non trivial Jacobian
\be \label{Jac}
det \left( \frac{\delta \varphi}{\delta \varphi'} \right) = 1 + i \int
d^4 x \, d^2 \theta \, \delta \alpha (x) \frac{1}{8} \frac{t_2 (A)}{8 \pi^2} 
W_a^{\alpha} W_{\alpha}^a.
\ee
where $t_2 (A)$ is the Dynkin index of the adjoint representation of $G$.\\
From Eq. (\ref{Jac}) one readily obtains the anomalous Ward-Takahashi identity
or ``Konishi equation''
\be
\left \langle \int \varphi_i^a (p) \frac{\delta S_{tot}}{\delta \varphi_i^a (p)
} + \frac{1}{8} \frac{t_2(A)}{8 \pi^2} \int W^{\alpha}_a W_{\alpha}^a \right 
\rangle = 0.
\ee
Now one can relate the anomalous term above to the singular quantity 
``$\delta_{ii}$'' (or its integral).\\
Since 
\be
\bea
{\ds 0 = \int d^4 p \, d^2 \theta \frac{1}{i} \frac{\delta}{\delta J_i (p)} 
\infu \frac{\delta}{\delta \varphi_i (p)} \exp \left(i S_{tot} \right) 
=}\nonumber\\
{\ds = \int d^4 p \, d^2 \theta \infu \frac{\delta}{\delta \varphi_i (p)} 
\left( \varphi_i (p) \exp \left(i S_{tot} \right) \right) }\\
\eea
\ee
one immediately obtains
\be \label{konano}
\int \sum_{i} \delta_{ii} = \left \langle \frac{i}{8}
\int d^2 \theta \, \frac{3 t_2(A)}{8 \pi^2} W_{\alpha}^a  
W^{\alpha}_a \right \rangle.
\ee
With this result, one can proceed with the derivation of ERG equation.\\
The RHS of Eq. (\ref{treq}) becomes
\be \label{treq2}
\bea
{\ds \int d^4 p \, \left( -\frac{1}{M} \right) \infu \left\{ \int d^2 \theta
\, \bigg[ -i M \delta_{ii} +M \varphi_i(p) \times \right. }\nonumber\\[0.3 cm]
{\ds \times \left. \frac{\delta S_{tot}} {\delta \varphi_i (p)} +
\frac{1}{2} \left( i \frac{\delta^2 S_{tot}}{\delta \varphi_i (p) 
\delta \varphi_i (-p)} - \frac{\delta S_{tot}}{\delta \varphi_i (p)}
\frac{\delta S_{tot}}{\delta \varphi_i (-p)} \right) \bigg ] + h.c. \right \} 
\exp \left(i S_{tot} \right). }\\
\eea
\ee
Separating $S_M$ term explicitly from $S_{tot}$, RHS becomes, after a little 
algebra,
\be \label{treq3}
\bea
{\ds \int d^4 p \, \infu \left \{ \int d^2 \theta \, \left[ \frac{i}{2} 
\delta_{ii} +\right. \right.}\nonumber\\[0.3 cm]
{\ds - \mdm \left( \frac{1}{2} M \varphi_i^2 \right) -\frac{1}{2 M} \left( 
i \frac{\delta^2 S_M}{\delta \varphi_i (p) 
\delta \varphi_i (-p)} - \frac{\delta S_M}{\delta \varphi_i (p)}
\frac{\delta S_M}{\delta \varphi_i (-p)} \right) +}\nonumber\\[0.3 cm]
{\ds \left. \left. -\frac{f^2(M)}{2 M} J_i(-p) J_i(p) -f(M) J_i(-p) \,
\varphi_i(p) \right] +h.c. \right \} \exp \left(i S_{tot} \right). }\\
\eea
\ee
Note that the term proportional to ${\ds i \delta_{ii} = - \frac{1}{8} \int 
\frac{3 t_2(A)}{8 \pi^2} W^2 }$
is contributed also by the double derivative term 
${\ds \frac{\delta^2}{\delta \varphi_i (p) \delta \varphi_i (-p)} \left( 
\frac{1}{2} M \int \varphi_i^2 \right)}$, part of 
${\ds \frac{\delta^2 
S_{tot}}{\delta \varphi_i (p) \delta \varphi_i (-p)} }$ term in Eq. (\ref{treq2}),
which explains the factor ${\ds \frac{1}{2} }$ in Eq. (\ref{treq3}).\\
From the inspection of Eq. (\ref{treq3}) one obtains the following conditions for
the $M$ invariance of low energy physics:
\begin{enumerate}
\item $S_M$ satisfies
\be \label{con1}
\bea
{\ds \frac{1}{2M} \int dp \, d^2 \theta \left( - \frac{\delta S_M}{\delta
\varphi_a^i(p)} \frac{\delta
S_M}{\delta \varphi_a^i(-p)} + i \frac{\delta^2 S_M}{\delta
\varphi_a^i(p) \delta \varphi_a^i(-p)} \right) +}\nonumber\\[0.38 cm]
{\ds + \frac{3}{16}
\int dp \, d^2 \theta \, \frac{t_2(A)}{8 \pi^2} W_a^{\alpha} W_{\alpha}^a
 + h.c. = \mdm S_M }\\
\eea
\ee
\item \be \label{con2}
\mdm f(M) = f(M) \quad {\it i.e.} \quad f(M) = \frac{M}{M_0}
\ee
\end{enumerate}
If the conditions Eq. (\ref{con1}) and Eq. (\ref{con2}) are satisfied then 
Eq. (\ref{treq3}) is seen to be equivalent to
\be
\infu \left[ - \mdm S_{tot} - \frac{f^2(M)}{2 M} \int J_i (-p) J_i(p) + h.c. 
\right] \exp \left(i S_{tot} \right)= 0
\ee
\ie
\be
\mdm Z_M + \left( i \frac{f^2(M)}{2 M} \int J_i (-p) J_i(p) + h.c. \right) Z_M = 0
\ee
\ie
\be \label{mdmZM}
\mdm \left[ \exp i \left( \frac{M}{2 M_0^2} \int J_i (-p) J_i(p) + h.c. \right)
Z_M \right] \doteq \mdm \tilde{Z}_M = 0.
\ee
The last equation implies that $Z_M$ is unchanged for varying value of $M$, except for the tree level two point function, \ie
\be \label{Z0tilde}
\tilde{Z}_0 \doteq \exp i \left( \frac{1}{2 M_0} \int J_i^2 + h.c. 
\right) Z_0 = \exp i \left( \frac{M}{2 M_0^2} \int J_i^2 + h.c. 
\right) Z_M = \tilde{Z}_M.
\ee
The flow equation (\ref{con1}) can be written as
\be \label{floweq}
\bea
{\ds \mdm \left \{ S_M - \frac{1}{16}
\int d^2 \theta \, \frac{3 t_2(A)}{8 \pi^2} W_a^{\alpha} W_{\alpha}^a 
\log\left( \frac{M}{M_0} \right) - h.c. \right \} =}\nonumber\\[0.3 cm]
{\ds = \frac{1}{2M} \int dp \, d^2 \theta \left( i \frac{\delta^2 S_M}{\delta
\varphi_a^i(p) \delta \varphi_a^i(-p)} - \frac{\delta S_M}{\delta
\varphi_a^i(p)} \frac{\delta S_M}{\delta \varphi_a^i(-p)} \right) +h.c. }\\
\eea
\ee
Since the anomalous term in Eq. (\ref{floweq}) does not depend on $\varphi_i$ or $
\bar{\varphi}_i$, one can express this result as
\be \label{floweq2}
\bea
{\ds \mdm \tilde{S}_M = \frac{1}{2M} \int dp \, d^2 \theta \left( i 
\frac{\delta^2 \tilde{S}_M}{\delta \varphi_a^i(p) \delta \varphi_a^i(-p)} - 
\frac{\delta \tilde{S}_M}{\delta \varphi_a^i(p)} 
\frac{\delta \tilde{S}_M}{\delta \varphi_a^i(-p)} \right) +h.c. }\nonumber\\[0.38 cm]
{\ds \tilde{S}_M \doteq S_M - \frac{1}{16}
\int d^2 \theta \, \frac{3 t_2(A)}{8 \pi^2} W_a^{\alpha} W_{\alpha}^a 
\log\left( \frac{M}{M_0} \right) - h.c. }\\
\eea
\ee
with the initial condition
\be
S_{M_0} = \tilde{S}_{M_0} = S_{N=4} (V, \varphi_i, \bar{\varphi}_i ; g_0).
\ee
\subsection{Change in holomorphic coupling constant.}
The low energy physics of $N=1$ SYM (at $p \sim M'<M<M_0$) is given by $S_M(V,
0,0)$.\\
The normal part $\tilde{S}_M (V,0,0)$ (see Eq. (\ref{floweq2})) is expected to 
vary rather slowly, \ie
\be
\tilde{S}_M (V,0,0) \simeq \tilde{S}_{M_0} (V,0,0) + {\cal O} \left(
\frac{1}{M} \right).
\ee
Then
\be
S_M (V,0,0) \simeq \frac{1}{16} \int d^4 p \, d^2 \theta \, 
\left( \frac{1}{g_0^2} +
\frac{3 t_2(A)}{8 \pi^2} \log\left( \frac{M}{M_0} \right) \right)
W_a^{\alpha} W_{\alpha}^a + h.c.
\ee
\ie, for $N=1$ SYM theory, the holomorphic gauge coupling constant changes according to
\be
\frac{1}{g^2(M_2)} - \frac{1}{g^2(M_1)} = \frac{3 t_2(A)}{8 \pi^2} 
\log\left( \frac{M_2}{M_1} \right)
\ee
or, in term of $\beta$ function
\be \label{hbeta}
\mdm \left( \frac{1}{g^2(M)} \right) = \frac{3 t_2(A)}{8 \pi^2}.
\ee
We refer to~\cite{mu1,mu2} for the detail of how to go over to the ``canonical
form'' from Eq. (\ref{hbeta}) and to obtain the ``exact'' expression of the $\beta$
function due to Novikov, Shifman, Vainshtein and Zakarov~\cite{NSV,SV}.\\ 
\section{The representation for $S_M$.}
\setcounter{equation}{0}
\label{sec:3}
\subsection{The solution of ERG equation.}
To have more accurate estimate of $\tilde{S}_M$ from Eq. (\ref{floweq2}), 
it would be convenient to have a representation of $S_M$ (or $\tilde{S}_M$) 
which makes its physical interpretation more explicit.\\
The formal solution of Eq. (\ref{floweq2}) with the given initial condition is
\be \label{solu}
\bea
{\ds \exp \left(i \tilde{S}_M (V,\varphi_i, \bar{\varphi}_i;g)
\right) = \exp \left\{ \frac{i}{2} \int \left(
\frac{1}{M_0} - \frac{1}{M} \right) \frac{\delta^2}{\delta \varphi_a^i(p) 
\delta \varphi_a^i(-p)} + h.c. \right\} \times}\nonumber\\
{\ds \times exp \Big(i S_{N=4} (V,\varphi_i,
\bar{\varphi}_i; g_0) \Big). }
\eea
\ee
If one applies the equality for the gaussian integral
\be
{\ds \exp \left( \frac{1}{2} \lambda^{-1}_{ij} \partial_{a_i} \partial_{a_j} 
\right) f(a) = \frac{{\ds \int_{{\bf R }^{N}} \prod_{i=1}^{N}  d x_i \, 
\exp \left(
-\frac{1}{2} \lambda_{ij} x_i x_j \right) f(x+a)}}{{\ds \int_{{\bf R}^{N}}
 \prod_{i=1}^{N} d x_i \, \exp \left(-\frac{1}{2} \lambda_{ij} x_i x_j 
\right)}} }
\ee
for positive definite $N \times N$ real matrix $\lambda$, then one can obtain 
the formally equivalent expression to Eq. (\ref{solu})
\be \label{intfor}
\bea
{\ds \exp \left(i \tilde{S}_M (V,\varphi_i, \bar{\varphi}_i) \right) = 
\Bigg \{ \int \prod_i {\cal D}[\varphi'_{i}] {\cal D}[\bar{\varphi}'_i] \exp i 
\bigg[ S_{N=4} (V, \varphi_i + \varphi'_{i}, \bar{\varphi}_i + 
\bar{\varphi}'_i;g_0) + }\nonumber\\[0.3 cm]
{\ds + \frac{\tilde{M}}{2} \int \varphi_{i}^{'2} + h.c. 
\bigg ] \Bigg \} \Bigg \{ \int \prod_i {\cal D}[\varphi'_{i}] 
{\cal D}[\bar{\varphi}'_i] 
\exp i \bigg[ \frac{\tilde{M}}{2} \int \varphi_{i}^{'2} + h.c. \bigg] 
\Bigg \}^{-1} }\\
\eea
\ee
where the ``reduced'' mass $\tilde{M}$ is defined by $\tilde{M}^{-1} \doteq 
M_0^{-1} - M^{-1}$.
The normalization factor in Eq. (\ref{intfor}) is important. In particular, in 
the limit $M \rightarrow M_0$, the numerator above should diverge as $| \delta
 M | \doteq |M-M_0| \rightarrow 0$.\\
Indeed, by rescaling anomaly
\be
\bea
{\ds \lim_{M \rightarrow M_0} \int \prod_i {\cal D}[\varphi'_{i}] 
{\cal D}[\bar{\varphi}'_i] \exp i \left\{S_{N=4} (V, \varphi_i + 
\varphi'_{i}, 
\bar{\varphi}_i + \bar{\varphi}'_i;g_0) + \frac{\tilde{M}}{2} \int 
\varphi_{i}^{'2} + h.c. \right \} \sim }\nonumber\\[0.38 cm]
{\ds \sim \exp i \left\{S_{N=4} (V, \varphi_i,\bar{\varphi}_i;g_0) + 
\frac{1}{16} \int \frac{3 t_2(A)}{8 \pi^2} W^2 \log \left( \frac{| \delta M |}
{M_0} \right) + h.c. \right\}. }\\
\eea
\ee
\subsection{Zinn Justin'~s transformation.}
To obtain Eq. (\ref{intfor}) in more rigorous fashion, one applies the 
transformation introduced in~\cite{MB}. Note that in what follows the
gaussian integral of the type
\be
f(\mu;J_i) \doteq \inffu \exp i \left \{ \frac{\mu}{2} \int \varphi_i^2 + h.c.
 + \int J_i \varphi_i +h.c. \right \}
\ee
inherently depends on the external gauge superfield $V$ (or $W$). One can 
evaluate such an ``anomalous'' gaussian integral by interpreting $f(\mu;J_i)
$ as
\be
\bea
{\ds f(\mu;J_i) = \lim_{\epsilon \rightarrow 0} f_{\epsilon} (\mu;J_i) }
\nonumber\\    
{\ds f_{\epsilon} (\mu;J_i) \doteq \inffu \exp i \left\{ \epsilon \int d^4 
\theta \, \bar{\varphi}_i e^V \varphi_i +\frac{\mu}{2} \int \varphi_i^2 + h.c.
+ \int J_i \varphi_i +h.c. \right \}. }\\
\eea
\ee
From this definition, one can show that
\be \label{eq1}
\frac{f(\mu;J_i)}{f(\mu;0)} = \exp \left( \frac{-i}{2 \mu} \int J_i^2 + h.c. 
\right) \\
\ee
\be \label{anom}
\frac{f(\mu_2;0)}{f(\mu_1;0)} = \exp \left\{ \frac{-i}{16} \int \frac{3 t_2(A)}
{8 \pi^2} W^{\alpha}_a W_{\alpha}^a \log \left( \frac{\mu_2}{\mu_1} \right) + 
h.c. \right \}.
\ee
Eq. (\ref{anom}) can be trivially obtained if one applies the rescaling 
transformations $\varphi_i = \sqrt{\frac{\mu_0}{\mu_a}} \varphi'_i,$ 
$ a=1,2$ and makes use of Konishi anomaly (\ref{Jac}).\\
More convincingly, one can calculate the matrix element of the mass operator
following the method, {\it e.g.}, of~\cite{gates} (See appendix A).
Now let us consider the particular example of Eq. (\ref{eq1})
\be \label{partex}
\bea
{\ds \ifu \exp i \left[ \frac{1}{2} \int \left( M+\tilde{M} \right) 
\varphi^{'2}_i +h.c. + \int \varphi'_i \left( -\tilde{M} \varphi_i + 
\right. \right.}\nonumber\\[0.4 cm]   
{\ds + \left. \left. \frac{M}{M_0} J_i \right) + h.c. \right] \left\{ \ifu 
\exp i \left[ \frac{1}{2} \int \left( M+\tilde{M} \right) 
\varphi^{'2}_i +h.c. \right] \right\}^{-1} = }\nonumber\\[0.4 cm]
{\ds = \exp i \left[ \frac{1}{2} \int \left( M_0 - \tilde{M} \right) 
\varphi^2_i + h.c. + \int J_i \varphi_i + h.c. + \frac{1}{2} \int \left( 
\frac{1}{M_0} - \frac{M}{M_0^2} \right) J_i^2 + h.c. \right]. }
\eea
\ee
With the help of Eq. (\ref{partex}) some of the factors defining $Z_0$ can be
written as
\be \label{intex}
\bea
{\ds \exp i \left[ \frac{1}{2 M_0} \int J_i^2 + h.c. \right] \exp i \left[ 
\frac{M_0}{2} \varphi^2_i + h.c. + \int J_i \varphi_i + h.c. \right] =
 \exp i \left[ \frac{M}{2 M_0^2} \int J_i^2 + \right.}\nonumber\\[0.4 cm]
{\ds + \left. \frac{\tilde{M}}{2} \int \varphi^2_i
+ h.c. \right] \left \{ \ifu 
\exp i \left[ \frac{1}{2} \int \left( M+\tilde{M} \right) 
\varphi^{'2}_i + h.c. + \int \varphi'_i \times \right. \right. }
\nonumber\\[0.4 cm]
{\ds \times \left. \left. \left( -\tilde{M} \varphi_i + 
\frac{M}{M_0} J_i \right) + h.c. \right] \right\}\left\{ \ifu 
\exp i \left[ \frac{1}{2} \int \left( M+\tilde{M} \right) 
\varphi^{'2}_i +h.c. \right] \right\}^{-1}. } 
\eea
\ee
Substituting Eq. (\ref{intex}) into Eq. (\ref{Z0tilde}) and Eq. (\ref{Z0}), one obtains
\be
\bea
{\ds \tilde{Z}_0 (J_i, \bar{J}_i, J_V) = \exp i \left[ \frac{M}{2 M_0^2} \int 
J_i^2 + h.c. \right] \int {\cal D}[V] \left \{ \inffu \times \right. }\nonumber\\[0.4 cm]
{\ds \times \left. \ifu \exp i \bigg[ S_{N=4} (V, 
\varphi_i, \bar{\varphi}_i ; g_0) + \frac{M}{2} \int \varphi^{'2}_i + h.c. +
\right.}\nonumber\\[0.4 cm]
{\ds +\frac{\tilde{M}}{2}
\int \left( \varphi_i - \varphi'_i \right)^2 + h.c. + \frac{M}{M_0} \int J_i
\varphi'_i + h.c. + \int J_V V \bigg] \times}\nonumber\\[0.4 cm]
{\ds \times \left. \left \{ \ifu \exp i \left[ \frac{1}{2} \int \left( M+
\tilde{M} \right) \varphi^{'2}_i +h.c. \right] \right\}^{-1} \right\}.}\\
\eea
\ee
Defining new variables $\varphi^{''}_i \doteq \varphi_i - \varphi'_i$ and
assuming ${\cal D}[\varphi_i] \, {\cal D} [\bar{\varphi}_i] = {\cal D}
[\varphi^{''}_i] \, {\cal D} [\bar{\varphi}^{''}_i]$ the RHS becomes
\be
\bea
{\ds \exp i \left[ \frac{M}{2 M_0^2} \int J_i^2 + h.c. \right] 
\int {\cal D}[V] \left \{ \ifu \exp i \left[ \frac{M}{2} \int \varphi^{'2}_i +
h.c. +\right. \right. }\nonumber\\[0.4 cm]
{\ds + \left. \frac{M}{M_0} \int J_i \varphi'_i + h.c. + \int J_V V 
\right] \innffu \exp i \left[ S_{N=4} (V, \varphi'_i +\varphi^{''}_i, 
\bar{\varphi}'_i + \bar{\varphi}^{''}_i;g_0) + \right.}\nonumber\\[0.4 cm]
{\ds \left. +\frac{\tilde{M}}{2} \int 
\varphi_i^{''2} + h.c. \right] \left. \left \{ \ifu \exp i \left[ 
\frac{1}{2} \int \left( M+
\tilde{M} \right) \varphi^{'2}_i +h.c. \right] \right\}^{-1} \right\}.}\\ 
\eea
\ee
This is nothing but the definition of $\tilde{Z}_M$ (see Eq. (\ref{ZM}) and 
Eq. (\ref{Z0tilde})) if one identifies
\be \label{eq35}
\bea
{\ds \exp \left( i S_M (V,\varphi_i,\bar{\varphi}_i) \right) \equiv \ifu 
\exp i \left[ S_{N=4} (V, \varphi_i + \varphi'_i, 
\bar{\varphi}_i + \bar{\varphi}'_i;g_0) +\right.}\nonumber\\[0.4 cm]  
{\ds + \left. \frac{\tilde{M}}{2} \int 
\varphi_i^{'2} + h.c. \right] 
\left \{ \ifu \exp i \left[ \frac{1}{2} \int \left( M+
\tilde{M} \right) \varphi^{'2}_i +h.c. \right] \right\}^{-1}. }
\eea
\ee
Eq. (\ref{eq35}) differs from Eq. (\ref{intfor}) only by the normalization 
factor. We rewrite, therefore, the RHS as
\be \label{intfor2}
\bea
\frac{{\ds \ifu \exp i \left[ \frac{\tilde{M}}{2} \int \varphi^{'2}_i +h.c. 
\right] }}{{\ds \ifu \exp i \left[ \frac{1}{2} \int \left( M+
\tilde{M} \right) \varphi^{'2}_i +h.c. \right] } }\times \nonumber\\
\times \frac{{\ds \ifu 
\exp i \left[ S_{N=4} (V, \varphi_i + \varphi'_i, 
\bar{\varphi}_i + \bar{\varphi}'_i;g_0) + \frac{\tilde{M}}{2} \int 
\varphi_i^{'2} + h.c. \right] }}{{\ds \ifu \exp i \left[ \frac{\tilde{M}}{2} \int 
\varphi^{'2}_i +h.c. \right] }}.  \\
\eea
\ee
The first factor is equal, according to Eq. (\ref{anom}), to
\be
\exp \left[ \frac{i}{16} \int \frac{3 t_2(A)}
{8 \pi^2} W^{\alpha}_a W_{\alpha}^a \log \left( \frac{M+\tilde{M}}{\tilde{M}} 
\right) + h.c. \right].
\ee
But
\be
\frac{M+\tilde{M}}{\tilde{M}} = 1 +\frac{M}{\tilde{M}} = 1 + M \left( 
\frac{1}{M_0} - \frac{1}{M} \right) = \frac{M}{M_0}. 
\ee
Thus Eq. (\ref{intfor2}) is equivalent to
\be \label{intfor3}
\bea
{\ds \exp \left( i S_M (V,\varphi_i,\bar{\varphi}_i) \right) = 
\exp \left[ \frac{i}{16} \int \frac{3 t_2(A)}
{8 \pi^2} W^{\alpha}_a W_{\alpha}^a \log \left( \frac{M}{M_0} \right) + 
h.c. \right] \times}\nonumber\\[0.3 cm]
{\ds \times \ifu \exp i \left[ S_{N=4} (V, \varphi_i + \varphi'_i, 
\bar{\varphi}_i + \bar{\varphi}'_i;g_0) + \frac{\tilde{M}}{2} \int 
\varphi_i^{'2} + h.c. \right] \times}\nonumber\\
{\ds \times \left\{ \ifu \exp i \left[ \frac{\tilde{M}}{2} \int 
\varphi^{'2}_i +h.c. \right] \right\}^{-1} }.
\eea
\ee
We have recovered Eq. (\ref{floweq2}) with $\tilde{S}_M$ given indeed by Eq. (\ref{intfor})!\\
Note that Eq. (\ref{floweq2}) or, equivalently, Eq. (\ref{intfor3}) is exact 
and neither depends on perturbation expansion nor it receives ``order 
$\frac{1}{M}$'' correction as the authors of~\cite{mu1,mu2} appear to hint.\\
Eq. (\ref{intfor3}) also shows the relationship of Wilsonian action
$S_M$ to generating functional of connected part $W (J_i,\bar{J}_i,J_V; 
\tilde{M})$. Such a relationship is well known in the ERG method with 
conventional cut-off (See~\cite{MB}).\\ 
\section{Conclusions.} 
Our main result is Eq. (\ref{floweq2}), which shows clearly the ``anomaly 
origin'' of Shifman-Veinstein result. We agree completely with the conclusions
of ~\cite{mu1,mu2}, but give a more precise definition of the ``anomalous''
and ``normal'' part of the Wilsonian action. While in the above refs the 
predominance of the ``anomalous'' part is justly emphasized, we 
put the residual (normal) part on the exact footing, writing the
equation (Polchinski'~s) obeyed by it.\\
Note that the remark about the relevance of ``vacuum energy'' is not limited
to the particular (and peculiar) regularization scheme which we have adopted.
If one had applied the usual cut-off method~\cite{Po} with or without some
prescriptions to guarantee the gauge invariance, one would have ended up with
similar anomalous terms through the vector field rescaling anomaly (really
the rescaling anomaly of accompanying F-P ghosts. See Ref.~\cite{mu2}).\\
This also raises the question about the non supersymmetric theory such as the
gauge coupling of left handed fermions, where the chiral anomaly~\cite{BOV} 
would appear like a rescaling anomaly.\\
These problems can be investigated with more confidence when the problem of
incorporating gauge invariance in ERG is solved.\\
Recently, T. Morris has proposed a novel analysis of the gauge 
problem in EGR method~\cite{Mor}. His prescription is still to be fully worked
out and at present rather cumbersome for the applications. However, the 
insights which this method appears to confer make us hope that we are getting
near the right solution at last.\\
It would be very interesting to study the significance of ``vacuum energy 
term'' in this new approach.\\
\section*{Acknowledgements.}
The authors wish to thank their colleagues for useful discussions as well for encouragement, in particular M. Bianchi, M. Bonini, G. C. Rossi, F. Vian and
F. Zwirner.\\
K. Y. wishes to thank K. Fujikawa and H. Kawai for constructive comments. He 
also wishes to thank Prof. H. Sugawara for the hospitality at KEK, Tsukuba
where part of the work has been carried out.\\ 

\appendix

\section{The evaluation of the gaussian integral.}
\setcounter{equation}{0}
Let
\be
\frac{{\ds \inffu \exp i \left[ \frac{\mu_2}{2} \int d^2 \theta \, \varphi^2_i
+ h.c. \right]}}{{\ds \inffu \exp i \left[ \frac{\mu_1}{2} \int d^2 \theta \, 
\varphi^2_i + h.c. \right]}} = \frac{{\ds f(\mu_2)}}{{\ds f(\mu_1)}}.
\ee
As we suggested in $\S$\ref{sec:3}, this should be interpreted as the limit
\be 
{\ds \lim_{\epsilon \rightarrow 0} \frac{{\ds \inffu \exp i \left[ 
\epsilon \int d^4 \theta \, \bar{\varphi}_i e^V \varphi_i + \frac{\mu_2}{2} 
\int d^2 \theta \, \varphi^2_i
+ h.c.\right]}}{{\ds \inffu \exp i \left[\epsilon \int d^4 \theta \, 
\bar{\varphi}_i 
e^V \varphi_i + \frac{\mu_1}{2} \int d^2 \theta \, \varphi^2_i + h.c. \right]
}} = \lim_{\epsilon \rightarrow 0} \frac{{\ds f_{\epsilon} (\mu_2)}}
{{\ds f_{\epsilon}(\mu_1)}}. }\\
\ee
Let us consider instead
\be
\bea
{\ds \left \langle \frac{\mu}{2} \int d^2 \theta \, \varphi^2_i + h.c. \right 
\rangle_{\epsilon} \doteq \inffu \left( \frac{\mu}{2} \int d^2 \theta \, 
\varphi^2_i + h.c. \right) \exp i \left[\epsilon \int d^4 \theta \, 
\bar{\varphi}_i e^V \varphi_i + \right.}\nonumber\\[0.2 cm]
{\ds \left. + \frac{\mu_1}{2} \int d^2 \theta \, \varphi^2_i + h.c. \right]
\left \{ \inffu \exp i \left[\epsilon \int d^4 \theta \, \bar{\varphi}_i 
e^V \varphi_i + \frac{\mu_1}{2} \int d^2 \theta \, \varphi^2_i + h.c. \right]
\right \}^{-1} =}\nonumber\\[0.2 cm]
{\ds = -i \mu \frac{\partial}{\partial \mu} f_{\epsilon} (\mu) \left[
f_{\epsilon} (\mu) \right]^{-1}. }\\
\eea
\ee
The matrix element can be explicitly evaluated with Feynman graphs 
(see~\cite{gates}).\\
The result is
\be
\left \langle \frac{\mu}{2} \int d^2 \theta \, \varphi^2_i + h.c. \right 
\rangle_{\epsilon} = - \frac{1}{16} \int \frac{3 t_2(A)}{8 \pi^2} W^2 + h.c. +
{\cal O} \left( \left( \frac{\epsilon}{\mu} \right)^4 \right).
\ee
Integrating the differential equation for $f_{\epsilon} (\mu)$ 
\be
\frac{{\ds f_{\epsilon} (\mu_2)}}{{\ds f_{\epsilon}(\mu_1)}} = \exp \left[ 
- \frac{i}{16} \int \frac{3 t_2(A)}{8 \pi^2} W^2 \log \left( 
\frac{\mu_2}{\mu_1} \right) + h.c. +
{\cal O} \left( \left( \frac{\epsilon}{\mu} \right)^4 \right) \right].
\ee
Taking the limit $\epsilon \rightarrow 0$
\be
\frac{{\ds f(\mu_2)}}{{\ds f(\mu_1)}} = \exp \left[ 
- \frac{i}{16} \int \frac{3 t_2(A)}{8 \pi^2} W^2 \log \left( 
\frac{\mu_2}{\mu_1} \right) + h.c. \right].
\ee
This is the result we wanted to prove.\\

\section{$\varphi^2$ insertion.}
\setcounter{equation}{0}
In this appendix\footnote{This calculation has been suggested by G. C. Rossi.}
, we consider the effect of introducing the gauge invariant 
source term
\be
\int d^4 x \, d^2 \theta \, K(x,\theta) \sum_{i=1}^{3} \varphi_i^2 (x,\theta)
\ee
instead of ${\ds \int J_i \varphi_i}$.\\
We consider the regularized $N=1$ SUSY Yang-Mills partition function (cf. Eq.
\ref{Z0})
\be \label{Z0primo}
\bea
{\ds Z'_0(J_i,\bar{J}_i,K,\bar{K},H,\bar{H},J_V) = \infu \exp i \left[ 
S_{N=4} (V,\varphi,\bar{\varphi}; g_0)+ \frac{M_0}{2} \times \right. }\nonumber\\
{\ds \times \int \varphi_i^2 
\left. + \int J_i \varphi_i + h.c. + \int J_V V + \frac{1}{2} \int
 K \varphi_i^2 + h.c. + \int H \frac{3}{16} \frac{t_2(A)}{8 \pi^2} W^2 
+ h.c. \right].}\\
\eea
\ee
With respect to the original definition Eq. (\ref{Z0}), we have added the
source terms for the composite (gauge invariant) operators $\varphi_i^2$, $W^2$
and their conjugates.\\
$J_i$, $\bar{J}_i$, $K$, $\bar{K}$, $H$, $\bar{H}$ and $J_V$ are space-time
dependent superfields.\\
Varying the regularizing mass $M_0$ to $M < M_0$, we would like to see if its
effect can be expressed by the simple modification of Eq. (\ref{ZM})
\be \label{ZMprimo}
\bea
{\ds Z'_M(\tilde{J}_i,\bar{\tilde{J}}_i,\tilde{K},\bar{\tilde{K}},\tilde{H},
\bar{\tilde{H}},J_V) = \infu \exp i \left[ \tilde{S}_M (V,\varphi,
\bar{\varphi}) + \frac{M}{2} \int  
\varphi_i^2 +\right. }\nonumber\\
{\ds \left. + h.c. + \int \tilde{J}_i \, \varphi_i + h.c. + \int J_V V +
\frac{1}{2} \int \tilde{K} \varphi_i^2 + h.c. + \int \tilde{H} \frac{3}{16} 
\frac{t_2(A)}{8 \pi^2} W^2 + h.c. \right]. }\\
\eea
\ee
In Eq. (\ref{ZMprimo}) one assumes that the ``Wilsonian action'' $\tilde{S}_M$
does not depend on the source terms    
$\tilde{J}_i$, $\bar{\tilde{J}}_i$, $\tilde{K}$, $\bar{\tilde{K}}$, 
$\tilde{H}$, $\bar{\tilde{H}}$ and $J_V$.\\
The new source superfields $\tilde{J}_i$, $\tilde{K}$, $\tilde{H}$ and their 
adjoints depend on the mass parameter $M$ and the original $J_i$, $K$ and $H$. 
They satisfy the initial conditions
\be
\bea
{\ds \tilde{J}_i (M=M_0) = J_i}\nonumber\\
{\ds \tilde{K} (M=M_0) = K}\nonumber\\
{\ds \tilde{H} (M=M_0) = H}
\eea
\ee
For the path integral of (\ref{ZMprimo}) the ``trivial'' equality Eq. 
(\ref{treq}) is still valid if one writes
\be
S_{tot} = \tilde{S}_M + \frac{M}{2} \int \varphi_i^2 + \int \tilde{J}_i
\, \varphi_i + h.c. + \int J_V V + \frac{1}{2}
\int \tilde{K} \varphi_i^2 + \int \tilde{H} \frac{3}{16} \frac{ 
t_2(A)}{8 \pi^2} W^2 + h.c.
\ee
As in $\S$\ref{sec:2}, we separate first the mass term as well as $\varphi_i$,
$\bar{\varphi}_i$ and $V$ source terms. Then Eq. (\ref{treq}) takes the form
\be 
\bea
{\ds 0 = \int d^4 p \, \frac{1}{M} \, \infu \Bigg \{ \int d^2 \theta \, \left[ 
- \frac{i}{2} \delta_{ii} +\right. }\nonumber\\[0.35 cm]
{\ds + \frac{1}{2} \left( 
i \frac{\delta^2 S^{ '}_M}{\delta \varphi_i (p) 
\delta \varphi_i (-p)} - \frac{\delta S^{ '}_M}{\delta \varphi_i (p)}
\frac{\delta S^{ '}_M}{\delta \varphi_i (-p)} \right) +
\frac{M^2}{2} \varphi_i(-p) \varphi_i(p) +}\nonumber\\[0.35 cm]
{\ds \left. + \frac{1}{2} \tilde{J}_i(-p) \tilde{J}_i(p) 
+ M \tilde{J}_i(-p) \,
\varphi_i(p) \right] +h.c. \Bigg \} \exp \left(i S_{tot} \right) }\\
\eea
\ee  
with ${\ds S^{ '}_M \doteq \tilde{S}_M + \frac{1}{2}
\int \tilde{K} \varphi_i^2 + h.c. + \int \tilde{H} \frac{3}{16} \frac{
t_2(A)}{8 \pi^2} W^2 + h.c. }$\\[0.3 cm]
Further separating $\tilde{S}_M$ from $S^{ '}_M$, it becomes
\be \label{treqprimo}
\bea
{\ds \infu \left \{ \frac{3}{16} \int \frac{t_2(A)}{8 \pi^2} W^2 \left ( 
\frac{M+\tilde{K}}{M} \right ) + \frac{1}{2 M} \int \left (M+\tilde{K} \right )^2 \varphi_i^2 + \right. }\nonumber\\[0.35 cm]
{\ds \left. + \frac{1}{2 M} \int \left( 
i \frac{\delta^2 \tilde{S}_M}{\delta \varphi_i (p) 
\delta \varphi_i (-p)} - \frac{\delta \tilde{S}_M}{\delta \varphi_i (p)}
\frac{\delta \tilde{S}_M}{\delta \varphi_i (-p)} \right) + \frac{1}{2 M} 
\int \tilde{J}_i(-p) \tilde{J}_i(p) + h.c. \right \} \exp \left (
i S_{tot} \right) = 0}\\
\eea
\ee
where we have replaced the ``anomaly'' ${\ds i \sum_i \delta_{ii} }$ by 
Konishi estimate $
{\ds - \frac{3}{8} \int \frac{t_2(A)}{8 \pi^2} W^2}$ 
(cf. Eq. \ref{konano}).\\
Note that the anomaly is now contributed also by the double derivative of the 
composite source term ${\ds \frac{\delta^2}{\delta \varphi_i^2 (x)} \frac{1}{2}
\int \tilde{K} \varphi_i^2(x) }$.\\
From Eq. (\ref{treqprimo}) one obtains the new conditions for $M$-invariance
\begin{enumerate}
\item   
\be \label{esse} 
\mdm \tilde{S}_M = \frac{1}{2M} \int \left( - \frac{\delta \tilde{S}_M}{\delta
\varphi_a^i(p)} \frac{\delta \tilde{S}_M}{\delta \varphi_a^i(-p)} + 
i \frac{\delta^2 \tilde{S}_M}{\delta \varphi_a^i(p) \delta \varphi_a^i(-p)} 
\right).
\ee
\item 
\be \label{acca}
\mdm \tilde{H} = \frac{\tilde{K} + M}{M}  \quad \mbox{with} \quad 
\tilde{H} (M=M_0) 
= H.
\ee
\item
\be \label{kappa}
\mdm \frac{1}{2} \left(\tilde{K} + M \right) = \frac{1}{2 M} \left(
\tilde{K} + M \right)^2 \quad \mbox{with} \quad \tilde{K} (M=M_0) = K.  
\ee
\item
\be \label{effe}
\bea
{\ds \tilde{J}_i = \tilde{f}(M) J_i }\nonumber\\
{\ds \mdm \tilde{f} = \frac{1}{M} \left(\tilde{K} + M \right) \tilde{f} 
\quad \mbox{with} \quad \tilde{f} (M=M_0) = 1. } 
\eea
\ee
\end{enumerate}
We see that the renormalized sources $\tilde{H}$ and $\tilde{J}_i$ depend on
$\tilde{K}$, the source of $\varphi_i^2$ term.\\
From the condition (\ref{kappa}) 
\be
\bea
{\ds \mdm \frac{1}{\tilde{K} + M} = \mdm \frac{1}{M} }\nonumber\\[0.3 cm]
{\ds \mbox{\ie} \quad \frac{1}{\tilde{K} + M} - \frac{1}{M} = const = \frac{1}{K + 
M_0} - \frac{1}{M_0}. }\\
\eea
\ee
Thus $\tilde{K}(x,\theta;M)$ depends on $M$ and the initial source factor 
$K(x,\theta)$ in ``non multiplicative'' way
\be \label{kappa2}
\frac{1}{\tilde{K}(x,\theta;M) + M} = \frac{1}{K(x,\theta) + M_0} - 
\frac{1}{\tilde{M}}.
\ee
From the condition (\ref{effe})
\be \label{effe2}
\bea
{\ds \partial_M \log \tilde{f} = \frac{1}{M^2} \left(\frac{1}{K + M_0} + 
\frac{1}{M} - \frac{1}{M_0} \right)^{-1} }\nonumber\\[0.3 cm]
{\ds \mbox{\ie} \quad \tilde{f}(M) = \frac{M}{M_0} \left[ 1 + \frac{M}{M_0} \left(
\frac{1}{M} - \frac{1}{M_0} \right) K \right]^{-1}. }\\
\eea
\ee
The condition (\ref{acca}) is equivalent to the equation for 
$\log \tilde{f}(M)$ above
\be \label{acca2}
\tilde{H}(x,\theta;M) = H(x,\theta) + \log \frac{M}{M_0} - \log \left[ 1 + 
\frac{M}{M_0} \left(\frac{1}{M} - \frac{1}{M_0} \right) K \right].
\ee
If the conditions (\ref{esse}), (\ref{kappa2}), (\ref{effe2}) and 
(\ref{acca2}) are satisfied, Eq. 
(\ref{treqprimo}) implies that the invariance equation analogous to Eq. (
\ref{mdmZM}) holds.
\be
\mdm \left \{ \exp i \left[ \frac{M}{2 M_0^2} \int \left( 1 + \frac{M}{M_0} 
\frac{K}{\tilde{M}} \right)^{-1} 
J_i (-p) J_i(p) + h.c. \right]
Z'_M \right \} \doteq \mdm \tilde{Z}'_M = 0.
\ee 
Again $Z'_M$ is unchanged except for the calculable tree level contribution. 
The extra factor disappears if one considers the case $J_i=\bar{J}_i=0$.\\
On the other hand, the results (\ref{kappa2}), (\ref{effe2}) and  (\ref{acca2})
 imply the mixing of operators, in particular $\varphi_i^2$ and $W^2$, when 
the regularizing mass $M$ is varied.\\
From Eqs (\ref{kappa2}), (\ref{effe2}) and (\ref{acca2})
\be
\bea
{\ds \frac{\partial \tilde{K}}{\partial K} = \left( \frac{M}{M_0} \right)^2 
\left( 1+\frac{M}{M_0} \frac{K}{\tilde{M}} \right)^{-2} }\nonumber\\[0.35 cm]
{\ds \frac{\partial \tilde{H}}{\partial K} = \frac{M}{M_0} \frac{1}{\tilde{M}} 
\left( 1+\frac{M}{M_0} \frac{K}{\tilde{M}} \right)^{-1} }\nonumber\\[0.35 cm]
{\ds \frac{\partial \tilde{F}}{\partial K} = \left( \frac{M}{M_0} \right)^2 
\frac{1}{\tilde{M}}
\left( 1+\frac{M}{M_0} \frac{K}{\tilde{M}} \right)^{-2}. }\\ 
\eea
\ee
If one takes the configurations with $J_i=\bar{J}_i=0$, then
\be
\bea
{\ds \left \langle \frac{1}{2} \sum_{i=1}^3 \varphi_i^2 \right \rangle_{M_0} =
\frac{1}{i Z_0} \frac{\delta Z_0}{\delta K} = \frac{1}{i Z_M} \left( 
\frac{\partial
 \tilde{K}}{\partial K} \frac{\delta}{\delta \tilde{K}} + \frac{\partial 
\tilde{H}} {\partial K} \frac{\delta}{\delta \tilde{H}} \right) Z_M =}
\nonumber\\
{\ds = \left( \frac{M}{M_0} \right)^2 
\left( 1+\frac{M}{M_0} \frac{K}{\tilde{M}} \right)^{-2} \left \langle 
\frac{1}{2} \sum_{i=1}^3 \varphi_i^2 \right \rangle_{M} + \frac{M}{M_0} 
\frac{1}{\tilde{M}} \left( 1+\frac{M}{M_0} \frac{K}{\tilde{M}} \right)^{-1} 
\left \langle \frac{3}{16} \frac{t_2(A)}{8 \pi^2} W^2 \right \rangle_{M} }
\eea
\ee
\ie
\be
\bea
{\ds \left \langle \frac{M_0}{2} \sum_{i=1}^3 \varphi_i^2 \right 
\rangle_{M_0} =
\frac{M}{M_0} \left( 1+\frac{M}{M_0} \frac{K}{\tilde{M}} \right)^{-2}  
\left \langle \frac{M}{2} \sum_{i=1}^3 \varphi_i^2 \right \rangle_{M} +
\left(1-\frac{M}{M_0} \right) \times }\nonumber\\[0.35 cm]
{\ds \times \left( 1+\frac{M}{M_0} \frac{K}{\tilde{M}} 
\right)^{-1} \left \langle - \frac{3}{16} \frac{t_2(A)}{8 \pi^2} W^2 
\right \rangle_{M}. }
\eea
\ee
On the other hand, from Eq. (\ref{acca2})
\be
\left \langle - \frac{3}{16} \frac{t_2(A)}{8 \pi^2} W^2 
\right \rangle_{M_0} =\left \langle - \frac{3}{16} \frac{t_2(A)}{8 \pi^2} 
W^2 \right \rangle_{M}.
\ee
For the ``physical'' configurations, \ie\ for $J_i = \bar{J}_i = K = \bar{K}
= H = \bar{H} = 0$ and arbitrary $J_V$  
\be
\left \langle \frac{M_0}{2} \sum_{i=1}^3 \varphi_i^2 \right \rangle_{M_0} =
\frac{M}{M_0} \left \langle \frac{M}{2} \sum_{i=1}^3 \varphi_i^2 \right 
\rangle_{M} + \left(1-\frac{M}{M_0} \right) \left \langle - \frac{1}{16} 
\frac{3 t_2(A)}{8 \pi^2} W^2 \right \rangle_{M} 
\ee
or, conversely,
\be
\left \langle \frac{M}{2} \sum_{i=1}^3 \varphi_i^2 \right \rangle_{M} =
\frac{M_0}{M} \left \langle \frac{M_0}{2} \sum_{i=1}^3 \varphi_i^2 \right 
\rangle_{M_0} + \left(1-\frac{M_0}{M} \right) \left \langle - \frac{1}{16} 
\frac{3 t_2(A)}{8 \pi^2} W^2 \right \rangle_{M_0}. 
\ee
 
\end{document}